\documentclass[12pt]{article}
\newcommand{\be}{\begin{equation}}
\newcommand{\ee}{\end{equation}}
\newcommand{\ba}{\begin{eqnarray}}
\newcommand{\ea}{\end{eqnarray}}

\begin{document}
\begin{center}
{\bf SUSY METHOD FOR THE THREE-DIMENSIONAL SCHR\"ODINGER EQUATION WITH EFFECTIVE MASS}\\
\vspace{0.5cm} {\large M. V. Iof\/fe$^{1,}$\footnote{E-mail: m.ioffe@spbu.ru, corresponding author},
E. V. Kolevatova$^{1,}$\footnote{E-mail: e.v.krup@yandex.ru},
D. N. Nishnianidze}$^{2,1,}$\footnote{E-mail: cutaisi@yahoo.com}\\
\vspace{0.5cm}
$^1$ Saint Petersburg State University, 7/9 Universitetskaya nab., St.Petersburg, 199034 Russia.\\
$^2$ Akaki Tsereteli State University, 4600 Kutaisi, Georgia.\\
\end{center}
\vspace{0.2cm} \hspace*{0.5in}
\vspace{1cm}
\hspace*{0.5in}
\begin{minipage}{5.0in}
{\small The three-dimensional Schr\"odinger equation with a position-dependent (effective) mass is studied in the framework of Supersymmetrical (SUSY) Quantum Mechanics.
The general solution of SUSY intertwining relations with first order supercharges is obtained without any preliminary constraints.
Several forms of coefficient functions of the supercharges are investigated and analytical expressions for the mass function and partner potentials are found.
As usual for SUSY Quantum Mechanics with nonsingular superpotentials, the spectra of intertwined Hamiltonians coincide up to zero modes of supercharges,
and the corresponding wave functions are connected by intertwining relations. All models are partially integrable by construction: each of them has at least one second order symmetry operator.} \\

\vspace*{0.1cm} PACS numbers: 03.65.-w
\end{minipage}


\vspace*{0.2cm}
\section{\bf Introduction.}
\vspace*{0.1cm} \hspace*{3ex}
Starting from the eighties of the last century, new method of study of different problems in nonrelativistic Quantum Mechanics was elaborated - the method of Supersymmetry \cite{witten}, \cite{reviews-1}-\cite{reviews-4}, \cite{ai}, which appeared as a reduction of the idea of Supersymmetry in Quantum Field Theory. Initially, this idea was designed to solve the various problems in Elementary Particle Physics, and the nonrelativistic one-dimensional Quantum Mechanics was chosen \cite{witten} as a suitable toy model for possible spontaneous breaking of Supersymmetry. Later on, the idea of Supersymmetry has been used very widely in different areas of relativistic physics, and Supersymmetrical Quantum Mechanics (SUSY QM) turned into a separate field for studying. It was applied both for investigation of qualitative analytical problems and for solution of some specific physical tasks. Among the first, multidimensional generalization \cite{abei-1}-\cite{abei-3}, Quantum Mechanics with matrix potential \cite{matrix-1}-\cite{matrix-8}, polynomial SUSY \cite{hsusy-1}-\cite{hsusy-11}, parasupersymmetry \cite{para-1}-\cite{para-3}, multiparticle systems \cite{calogero-1}-\cite{calogero-8} has to be mentioned. Among the second, the spectrum of Pauli fermions on a plane \cite{pauli-1}-\cite{pauli-6}, different variants of scattering problem \cite{baye-1}-\cite{baye-4}, quantum design \cite{ai}, \cite{design-1}-\cite{design-6}, nonlinear Ermakov-Milne-Pinney equation \cite{ermakov}, supersymmetrical WKB method \cite{WKB-1}, \cite{WKB-2} can be listed as examples.

One more direction of SUSY QM was not mentioned above - the study of non-relativistic quantum models with position-dependent mass (PDM). These models (also called as models with the effective mass) appear in many branches of physics very distinct from each other: quantum liquids \cite{-22}, nuclear physics \cite{nucl}, quantum wires and dots \cite{-25}, physics of semiconductors \cite{-21-1}-\cite{-21-4}, and some others. The SUSY QM approach was used successfully to investigate the wide class of such models - see for example \cite{quesne-1}, \cite{quesne-2}, \cite{CIN}, \cite{nikitin}, \cite{ioffe-PDM}. Although most of the papers on the effective mass were focused on one-dimensional systems, there were few two-dimensional \cite{quesne-1}, \cite{quesne-2}, \cite{CIN}, three-dimensional \cite{nikitin} and $(2+1)$-nonstationary \cite{schulze} exclusions. In particular, in \cite{CIN} the general solution of SUSY intertwining relations of first order in derivatives  was built for two-dimensional Schr\"odinger operators with position-dependent (effective) mass in terms of four arbitrary functions. The generalization for intertwining of second order in derivatives was also
considered in \cite{CIN}, and there some particular solutions were found. The paper \cite{quesne-1}, \cite{quesne-2} mainly deals with two-dimensional systems but with a mass function depending only on one coordinate. The paper \cite{nikitin} was devoted to the three-dimensional rotationally invariant PDM models starting from the condition of superintegrability. The complete set of such systems was built and classified. It was shown that these models possess the property of shape invariance, the important element of SUSY approach in Quantum Mechanics \cite{genden}, and admit exact solvability. In the present paper, just the three-dimensional Schr\"odinger equation with PDM will be studied in the frameworks of SUSY Quantum Mechanics without restriction to rotational invariance. Namely, the general solution of SUSY intertwining relations will be found.

The structure of the paper is the following. The SUSY intertwining relations are written out in Section 2 for the general pair of Hamiltonians with position dependent mass. Section 3 contains the starting steps of solution of intertwining relations in a most general form. Four different options of parameters values are considered separately in Section 4 completing the procedure of general solution of the problem. Some concluding remarks can be found in Conclusions.

\section{\bf SUSY intertwining relations for PDM Hamiltonians in $d=3$ space.}
\vspace*{0.1cm}
\hspace*{3ex}
First of all, we have to write down the Hermitian PDM Hamiltonian.
This problem is not quite trivial due to non-commutativity of the effective mass
function $m(\vec x)$ with three-vector of the momentum operator $\vec p=-i\vec{\partial}\equiv -i\partial/\partial{\vec x}. $
Different arguments exist for the choice of kinetic term (for example, see references in \cite{ioffe-PDM}), and we shall use the most
popular (explicitly Hermitian) form \cite{roos}:
\ba
  H&=&-\frac{1}{2}\biggl[M^{\alpha}(\vec x)\partial_iM^{\beta}(\vec x)\partial_iM^{\gamma}(\vec x) +
  M^{\gamma}(\vec x)\partial_iM^{\beta}(\vec x)\partial_iM^{\alpha}(\vec x)\biggr] + V(\vec x)=\nonumber\\
  &=&-\partial_i\frac{1}{M(\vec x)}\partial_i + U(\vec x),\label{roos}
\ea
where $\vec x\equiv (x_1,x_2,x_3),\,\,\partial_i\equiv \partial /\partial x_i,$ the effective mass function $m(\vec x)\equiv m_0M(\vec x),$ $\hbar = 2m_0=1,$ and
the summation over repeated indices will be implied. The effective
mass function $M(\vec x)$ is dimensionless function of coordinates, the real function $V(\vec x)$ is the physical potential, and
$\alpha , \beta , \gamma$ are constant parameters satisfying the only restriction $\alpha + \beta +\gamma = -1.$
These parameters are absorbed in the additional terms of effective potential
$U(\vec x):$
\begin{equation}
U(\vec x)= V(\vec x)+\frac{\beta +1}{2}\frac{\Delta^{(3)}M(\vec x)}{M^2(\vec x)}-
\biggl[\alpha(\alpha + \beta +1)+\beta +1\biggr]\frac{\biggl(\partial_iM(\vec x)\biggr)\biggl(\partial_iM(\vec x)\biggr)}{M^3(\vec x)};\quad
\Delta^{(3)}\equiv \partial_i\partial_i.
\label{UV}
\end{equation}
The specific values for $\alpha , \beta , \gamma$ will not be discussed in this paper, they depend on the specific physical
model under consideration.

The inclusion of the system into the framework of SUSY approach means that its Hamiltonian $H_1$ satisfies the following intertwining relations:
\ba
H_1 Q^+=Q^+H_2; \label{intertw1}\\
Q^-H_1=H_2Q^-, \label{intertw2}
\ea
where $H_2$ is the partner Hamiltonian for $H_1$, and $Q^{\pm}$ are the mutually Hermitian conjugate intertwining operators (off-diagonal components of matrix supercharges). Different realizations of these
relations exist: they depend both on the form of Hamiltonians (scalar, matrix, multidimensional) and on the order of differential operators $Q^{\pm}$ (linear or higher order in derivatives). Independently on realization, intertwining relations (\ref{intertw1}), (\ref{intertw2}) with nonsingular operators $Q^{\pm}$ lead to the isospectrality of Hamiltonians $H_{1,2},$ but up to possible zero modes of $Q^{\pm}$ (see \cite{isotonic} as example of model with singular superpotential). Correspondingly, the wave functions are obtained from each other (up to the constant multipliers):
\be
H_i\Psi^{(i)}_n(\vec x)=E_n\Psi_n^{(i)}(\vec x);\quad i=1,2;\quad n=0,1,2,...;\quad
\Psi_n^{(2)}=Q^-\Psi_n^{(1)};\quad \Psi_n^{(1)}=Q^+\Psi_n^{(2)}.
\label{schr}
\ee

Also, it should be noted that if the Hamiltonian participates in some intertwining relations, it automatically is partially integrable. The partial (Liouville)-integrability  of $d-$dimensional quantum system $d\geq 2$ means \cite{integr-1}, \cite{integr-2} that its Hamiltonian commutes with a number $k\leq (d-1)$ of mutually independent operators $R^{(l)},\, l=1,2,...,k$ (symmetry operators). The system is called completely integrable if $k=(d-1).$ In our $d=3$ case, due to
relations (\ref{intertw1}), (\ref{intertw2}), each Hamiltonian $H_{1}, H_2$ commutes with its symmetry operator $R_1=Q^+Q^-,\, R_2=Q^-Q^+,$ correspondingly:
\be
[H_i, R_i]=0;\quad i=1,2,
\label{symm}
\ee
i.e. $k<(d-1)$, and the system is at least partially integrable. It should be kept in mind that sometimes the symmetry operators $R_i$ may be expressed as function of the Hamiltonian $H_i$ itself, but for many systems this is not so (see \cite{david-1}-\cite{david-8}). In the present case, the part with highest derivatives in $R_1$ has the form
$-q_lq_m\partial_l\partial_m$ which can not be proportional to the highest derivative part $-1/M(\vec x)\partial_i\partial_i$ in $H_1$ due to the general expressions for $q_l$
(see Eqs.(\ref{Q+}) and (\ref{1-4}) - (\ref{1-6}) below). Therefore, the full operators $R_1$ and $H_1$ can not be proportional to each other, as well. The same conclusion is true for $R_2$ and $H_2.$

Hereafter, the three-dimensional Hamiltonians $H_1,\, H_2$ in (\ref{intertw1}), (\ref{intertw2}) are of the form (\ref{roos})
with the same effective mass $M(\vec x),$ but with different effective potentials $U_{1,2}(\vec x)$ and, correspondingly, different potentials $V_{1,2}(\vec x)$ in (\ref{UV}). The intertwining operators are chosen of the most general first order form:
\begin{equation}\label{Q+}
  Q^+=(Q^-)^{\dagger}=q_l(\vec x)\partial_l+q(\vec x); \quad l=1,2,3 .
\end{equation}

It is convenient to perform the similarity transformation which allows to eliminate the first derivatives from the kinetic term in (\ref{UV}) (the same method was used in \cite{CIN}):
\ba
h_{1,2}&=&e^{-\phi (\vec x)}H_{1,2} e^{\phi (\vec x)}\equiv -\frac{1}{M(\vec x)}\Delta^{(3)}+v_{1,2}(\vec x);
\label{similarity}\\
v_{1,2}(\vec x)&=&U_{1,2}(\vec x)-\frac{\Delta^{(3)}M(\vec x)}{2M^2(\vec x)}+
\frac{3\biggl( \partial_kM(\vec x) \biggr)\biggl( \partial_kM(\vec x)\biggr)}{4M^3(\vec x)}. \label{vU}
\ea
The result is:
\begin{equation}\label{similarity2}
  q^{+}=e^{-\phi (\vec x)}Q^{+}e^{\phi (\vec x)}= q_l(\vec x)\partial_l+p(\vec x);\quad
p(\vec x)=q(\vec x)+\frac{q_k(\vec x)\biggl( \partial_kM(\vec x) \biggr)}{M(\vec x)}.
\end{equation}
The condition of absence of first order derivatives in (\ref{similarity}) reads:
\be
\partial_k\phi (\vec x)=\frac{\biggl( \partial_kM(\vec x)\biggr)}{2M(\vec x)};\quad
\phi (\vec x)=\frac{1}{2}\ln M(\vec x) + Const.
\label{phi}
\ee
Thus, the function $\phi$ is real, the similarity transformation (\ref{similarity}), (\ref{similarity2}) is not unitary, and therefore,
$h_{1,2}$ are not Hermitian, and $q^+\neq (q^-)^{\dagger}.$ Nevertheless, these
operators satisfy the intertwining relation equivalent to (\ref{intertw1}):
\be
h_1 q^+=q^+h_2.
\label{intertw}
\ee
In other words, each solution of (\ref{intertw1}) gives the only solution of (\ref{intertw}), and vice versa. More of that,
(\ref{intertw2}) will be automatically fulfilled, being the Hermitian conjugate of (\ref{intertw1}).

Now, we shall deal with (\ref{intertw}), which seems to be much simpler than the initial (\ref{intertw1}).
The relation (\ref{intertw}) is equivalent to a system of ten nonlinear differential equations for the functions $M,\, v_{1,2},\,q_i,\, p:$
\ba
&&M(\vec x)\biggl[ \biggl( \partial_kq_i(\vec x)\biggr)+\biggl(\partial_iq_k(\vec x) \biggr) \biggr] +
\delta_{ik}q_j(\vec x)\biggl( \partial_jM(\vec x) \biggr)=0;
\label{1}\\
&&2M(\vec x)v(\vec x)q_i(\vec x)-2\biggl( \partial_ip(\vec x)\biggr)-\biggl( \Delta^{(3)}q_i(\vec x)\biggr)=0;
\label{2}\\
&&M(\vec x)\biggl[ q_i(\vec x)\biggl( \partial_iv_2(\vec x) \biggr)-2v(\vec x)p(\vec x)\biggr]+
\biggl( \Delta^{(3)}p(\vec x)\biggr)=0,
\label{3}
\ea
where indices take values $1,2,3$, and the function $v(\vec x)$ is defined as:
\be
2v(\vec x)\equiv v_1(\vec x)-v_2(\vec x). \label{vvv}
\ee

\section{\bf Solution of intertwining relations.}
\vspace*{0.1cm}
\hspace*{3ex}
We begin from the solution of the first three equations (\ref{1}). They provide the expressions for functions $q_i(\vec x):$
\ba
q_1(\vec x)&=&b(x_1^2-x_2^2-x_3^2)+2ax_1x_3+2cx_1x_2+ex_1+kx_2-nx_3+e_1;\label{1-4}\\
q_2(\vec x)&=&c(x_2^2-x_1^2-x_3^2)+2ax_2x_3+2bx_1x_2+ex_2-kx_1+mx_3+e_2;\label{1-5}\\
q_3(\vec x)&=&a(x_3^2-x_1^2-x_2^2)+2cx_2x_3+2bx_1x_3+ex_3+nx_1-mx_2+e_3,\label{1-6}
\ea
and also the equation for the mass function $M(\vec x):$
\be
q_j(\vec x)\partial_jM(\vec x)=-2M(\vec x)\partial_1q_1(\vec x).\label{MM}
\ee
The system (\ref{2}) for functions $q_i(\vec x)$ above can be rewritten as follows:
\be
M(\vec x)v(\vec x)q_1+b=\partial_1p(\vec x);\quad M(\vec x)v(\vec x)q_2+c=\partial_2p(\vec x);\quad M(\vec x)v(\vec x)q_3+a=\partial_3p(\vec x). \label{22}
\ee
From the system
Eq.(\ref{22}) one can derive the following relation for functions $q_i:$
\ba
&&A\equiv q_1\partial_2q_3 + q_2\partial_3q_1+q_3\partial_1q_2= -(mb+nc+ka)x_ix_i+x_1(-me+2ae_2-2ce_3)+\nonumber\\
&&x_2(-ne+2be_3-2ae_1)+x_3(-ke+2ce_1-2be_2)-(me_1+ne_2+ke_3)=0.\label{3-2}
\ea
From (\ref{3}) using (\ref{2}) and (\ref{MM}), one obtains equation for $p(\vec x)$:
$$
\Delta^{(3)}p(\vec x)=\partial_i(Mvq_i)=M(q_i\partial_iv+v\partial_1q_1) \nonumber
$$
where summation over repeated indices $i=1,2,3$ was assumed, and the relation $\partial_1q_1=\partial_2q_2=\partial_3q_3$ was explored.
The substitution of the expression for $\Delta^{(3)}p(\vec x)$ into (\ref{3}) allows to eliminate the mass $M(\vec x)$ from (\ref{3}), providing an additional relation between unknown functions $p, v_2, v$:
\be
q_i\partial_i(v_2+v)=v(2p-\partial_1q_1) \label{1-9}
\ee
It is necessary to consider separately different particular values of constants.

\section{\bf Particular cases of parameter values.}
\vspace*{0.1cm}
\hspace*{3ex}
The procedure of calculation of the mass function and potentials depends essentially on the number of vanishing parameters $a, b, c,$ i.e. on
the form of quadratic part of functions $q_i$ in (\ref{1-4}) - (\ref{1-6}).

{\bf Case I.} $\qquad \qquad \qquad a=b=c=0.$\\
In this case, all $q_i$ are linear functions:
\be
q_1=ex_1+kx_2-nx_3+e_1; \quad q_2=-kx_1+ex_2+mx_3+e_2;\quad q_3=nx_1-mx_2+ex_3+e_3.\label{q3}
\ee
One can try to exclude the free terms $e_i$ by means of suitable shift of coordinates $x_i\to x_i+\Delta_i$ with constants $\Delta_i.$ This opportunity depends on
the determinant of $3\times 3$ matrix of coefficients in (\ref{q3}) which equals $e(e^2+m^2+k^2+n^2).$

{\bf Case Ia.} $\qquad\qquad\qquad  e\neq 0.$\\
Now, the constant shifts $\Delta_i$ can be calculated, and actually all $e_i$ can be taken $e_i=0.$ Then, (\ref{3-2}) gives $m=n=k=0,$ and (\ref{q3})
leads to $q_i=ex_i.$

For such $q_i,$ Eqs.(\ref{22}) in the spherical coordinates $r, \theta, \varphi $ provide:
$$
(x_2\partial_1-x_1\partial_2)p(\vec x)=\partial_{\varphi}p(r, \theta, \varphi)=0; \,\,
(x_3\partial_1-x_1\partial_3)p(\vec x)=(\cos\varphi\partial_{\theta}-\cot\theta\sin\varphi\partial_{\varphi})p(r, \theta, \varphi)=0.  \nonumber
$$
Therefore, the function $p$ depends only on the radial variable $p(\vec x)=p(r),$
and its derivative over $x_i$ is:
$$
\partial_ip(r)=\frac{x_i}{r}p^{\prime}(r). \nonumber
$$
Using again (\ref{22}), one obtains:
\be
M(\vec x)v(\vec x)=\frac{p^{\prime}(r)}{er}. \label{b}
\ee

In its turn, Eq.(\ref{MM}) provides the form of mass function:
\be
2eM(r, \theta, \varphi)=-er\partial_rM(r, \theta, \varphi),\quad M(r, \theta, \varphi)=\frac{N(\theta, \varphi)}{r^2}, \label{7-0}
\ee
where $N(\theta, \varphi)$ is an arbitrary function of angles.
This equation together with (\ref{b}) leads to the function $v(\vec x):$
\be
v(\vec x)=\frac{rp^{\prime}(r)}{eN(\theta, \varphi)}. \label{c}
\ee

Now, the expression for potential $v_2(r, \theta, \varphi)$ can be calculated by integration of Eq.(\ref{1-9}) written in a form:
$$
er\partial_r(v_2+v)=v(2p-e), \nonumber
$$
and using (\ref{7-0}), (\ref{c}):
$$
er\partial_r(v_2+v)=\frac{rp^{\prime}(r)}{eN(\theta, \varphi)}(2p(r)-e). \nonumber
$$
Thus,
\be
v_{1,2}(r, \theta, \varphi)=\frac{1}{e^2N(\theta, \varphi)}\biggl( \pm erp^{\prime}(r)+p^2(r)-ep(r)\biggr) + L(\theta, \varphi), \label{7-00}
\ee
with a second arbitrary angle function $L(\theta, \varphi).$

{\bf Case Ib.} $\qquad\qquad   e=0.$\\
For such $e,$ two opportunities are open. The first one,\\
{\bf Case Ib.1.}  $\qquad\qquad q_1=kx_2-nx_3;\, q_2=-kx_1+mx_3;\, q_3=nx_1-mx_2,$\\
so that the linear combination vanishes:
\be
mq_1+nq_2+kq_3=0. \label{8-1}
\ee
New variables are convenient here:
\be
y_1=mx_1+nx_2+kx_3;\,\,\, y_2=q_2=-kx_1+mx_3;\,\,\, y_3=q_3=nx_1-mx_2;\,\,\, m\neq 0. \nonumber
\ee
The Eq.(\ref{MM}) can be rewritten as:
\be
\biggl[ \biggl((m^2+k^2)y_3+kny_2\biggr)\partial_{y_2}-\biggl((n^2+m^2)y_2+kny_3\biggr)\partial_{y_3} \biggr]M=0, \nonumber
\ee
and taking into account
\be
q_i\partial_i=\frac{1}{m}\biggl[ \biggl((m^2+k^2)y_3+kny_2\biggr)\partial_{y_2}-\biggl((n^2+m^2)y_2+kny_3\biggr)\partial_{y_3} \biggr];\,\, \partial_1q_1=0, \label{8-3}
\ee
Eq.(\ref{MM}) can be integrated explicitly:
\be
M=M\biggl(y_1, (m^2+n^2)y_2^2+(m^2+k^2)y_3^2+2kny_2y_3 \biggr)    \label{9-1}
\ee
with $M$ - an arbitrary function of its two arguments. From the system (\ref{22}), it follows due to (\ref{8-1}) that the function $p$
actually depends only on two variables $y_2, y_3$:
$$
(mq_1+nq_2+kq_3)Mv=(m^2+k^2+n^2)\partial_{y_1}p=0, \quad i.e. \quad p=P(y_2, y_3). \nonumber
$$
The Eq.(\ref{22}) is:
$$
Mvy_2=-m\partial_{y_3}P;\quad Mvy_3=m\partial_{y_2}P,  \nonumber
$$
and therefore it gives new restrictions onto the function $P$ and the combination $Mv$:
$$
P(y_2, y_3)=P(\frac{y_2}{y_3});\quad\quad Mv=\frac{m}{y_3^2}P^{\prime}(\frac{y_2}{y_3}). \nonumber
$$
The last equation we must consider in the case Ib.1 is (\ref{1-9}). Taking into account that $q_i\partial_iM=0,$ one obtains:
\be
q_i\biggl(\partial_iM(v_2+v)\biggr)=2MvP=\frac{m}{y_3^2}\biggl(\partial_{z_2}P^2\biggr),\quad z_2\equiv \frac{y_2}{y_3},   \label{10-1}
\ee
where we introduced new variables:
\be
z_1=\frac{1}{2}(m^2+n^2)y_2^2+\frac{1}{2}(m^2+k^2)y_3^2+kny_2y_3;\,\, z_2=\frac{y_2}{y_3};\,\, z_3=y_1, \label{10-2}
\ee
such that (\ref{8-3}) reads:
\be
q_i\partial_i=\frac{1}{m}\biggl( m^2+k^2+2knz_2+(m^2+n^2)z_2^2 \biggr)\partial_{z_2}. \nonumber
\ee
Eqs.(\ref{10-1}), (\ref{10-2}), where the function (\ref{9-1}) is considered now as a function of new variables (\ref{10-2}) $M=M(z_3, z_1),$ provide:
$$
\frac{2z_1M}{m}\biggl(\partial_{z_2}(v_2+v)\biggr)=m\biggl(\partial_{z_2}P^2(z_2)\biggr), \nonumber
$$
so that
$$
v_2+v=\frac{m^2P^2(z_2)}{z_1M(z_3, z_1)}+N(z_3, z_1), \nonumber
$$
with an arbitrary function $N(z_3, z_1).$ Thus, the partner potentials are:
\be
v_{1,2}=\frac{m}{z_1M(z_3, z_1)}\biggl[mP^2(z_2)\pm \biggl((m^2+n^2)z_2^2+2knz_2+m^2+k^2 \biggr)P^{\prime}(z_2) \biggr]+N(z_3, z_1). \label{a0}
\ee

The second option of Case Ib,\\
{\bf Case Ib.2} $\qquad\qquad q_i=e_i.$\\
Suitable variables for this option are:
$$
y_1=e_ix_i;\quad y_2=e_2x_1-e_1x_2;\quad y_3=e_3x_1-e_1x_3;\quad e_1\neq 0, \nonumber
$$
and $e_i\partial_i=e_ie_i\partial_{y_1}.$ From (\ref{MM}), one obtains that $e_i\partial_iM(\vec x)=0$ and therefore $M=M(y_2, y_3).$ Relations (\ref{22})
are rewritten as:
$$
(e_3\partial_2-e_2\partial_3)p(\vec x)=(e_1\partial_3-e_3\partial_1)p(\vec x)=0 \nonumber
$$
and in terms of $y_i$ are equivalent to:
$$
(e_2\partial_{y_3}-e_3\partial_{y_2})P(\vec y)=0;\quad \biggl[ (e_1^2+e_3^2)\partial_{y_3}+e_2e_3\partial_{y_2}\biggr]P(\vec y)=0, \nonumber
$$
where $p(\vec x)\equiv P(\vec y).$
Thus, $\partial_{y_2}P=\partial_{y_3}P=0$, i.e. $P$ depends only on one variable $P=P(y_1),$ and $Mv=P^{\prime}(y_1).$ Eq.(\ref{1-9}) provides:
$$
v_2+v=\frac{P^2(y_1)}{(e_ie_i)M}+S(y_2, y_3) \nonumber
$$
with an arbitrary function $S(y_2,y_3).$ Correspondingly, the partner potentials are:
\be
v_{1,2}=\frac{1}{M(y_2, y_3)}\biggl[ \frac{P^2(y_1)}{(e_ie_i)}\pm P^{\prime}(y_1)\biggr]+S(y_2, y_3). \label{b0}
\ee

In order to illustrate the Case Ib.2 by some specific example, let us consider the following particular model. For simplicity, we shall take $e_2=e_3=0;\, e_1=1,$ i.e. $y_1=x_1,\, y_2=-x_2, \, y_3=-x_3,$
and $q^{\pm}=\pm \partial_1+P(x_1)=Q^{\pm}.$ Also, the second term in the potentials (\ref{b0}) will be taken $S(y_2,y_3)=0,$ and the mass function $M(y_2,y_3)$ will be chosen rotationally
invariant $M(y_2,y_3)\equiv 1/\rho$ with $\rho\equiv \sqrt{x_2^2+x_3^2}.$ The equation $h_1\psi(\vec x)=E^{(1)}\psi(\vec x)$ is amenable to separation in terms of
cylindrical variables $x_1,\,\rho, \phi ,$ since
$$h_1=-\frac{1}{M(\rho)}\Delta^{(3)}+v_1(x_1,\rho,\phi)=\frac{1}{M(\rho)}\biggl[-\partial_{x_1}^2-\partial_{\rho}^2-\frac{1}{\rho}\partial_{\rho}-\frac{1}{\rho^2}\partial_{\phi}^2+
\biggl(P^2(x_1)+P'(x_1)\biggr)\biggr].$$
The eigenfunctions of $h_1$ are the linear combinations of factorized terms $\psi_{x_1}(x_1)\psi_{\rho}(\rho)\psi_{\phi}(\phi).$ Periodicity of wave functions in $\phi$ provides that $\partial_{\phi}^2 \sim -n^2;\, n=0,\pm 1,\pm 2... ,$ and the part depending on $x_1$ has the form of one-dimensional Schr\"odinger equation
$\biggl[-\partial_{x_1}^2+\biggl(P^2(x_1)+P'(x_1)\biggr)\biggr]\psi_l (x_1)=\epsilon_l\psi_l(x_1)$ with eigenvalues $\epsilon_l; \, l=0,1,2,... .$ Therefore, one has to find the spectrum of the operator $h_1$ by studying the radial equation:
$$
\biggl[\rho^2\partial_{\rho}^2+\rho\partial_{\rho}-\epsilon_l\rho^2+E^{(1)}\rho-n^2\biggr]\psi_{\rho}(\rho)=0.
$$
After the suitable transformation of variables and functions, $\psi_{\rho}(\rho)\equiv \rho^n u(\rho);\,\, u(\rho)\equiv e^{-x/2}y(x),$ where $x\equiv 2\sqrt{\epsilon_l}\rho ;\,\, \epsilon_l >0,$ the eigenfunctions $\psi_{\rho;n,l}(\rho)$ and the corresponding energy eigenvalues $E^{(1)}_{n,l}$  depend on quantum numbers $n,\,l.$ The eigenfunctions
have the form $\psi_{\rho;n,l}(\rho)=\rho^ne^{-\sqrt{\epsilon_l}\rho}y(2\sqrt{\epsilon_l}\rho),$ where $y(x)$ is a solution $\Phi(a,c;x)$ of Confluent Hypergeometric Equation \cite{bateman} with parameters
$a=\frac{1}{2}(1+2n-E^{(1)}/\sqrt{\epsilon_l});\, c=(1+2n).$ The condition of normalizability of the wave functions $\Psi^{(1)}$ for the Hamiltonian $H_1$ (see (\ref{similarity})) provides the discreet energy spectrum of the model: $E^{(1)}_{n,l,m}=(1+2n+2m)\sqrt{\epsilon_l},$ where $m$ is the new positive integer quantum number. The
corresponding wave functions are $\Psi^{(1)}_{n,l,m}=e^{\pm in\phi}\rho^{n-1/2}e^{-\sqrt{\epsilon_l}\rho}y(2\sqrt{\epsilon_l}\rho)\psi_{x_1}(x_1),$ where $y(x)=L^{(2n)}_m(x)$
are the generalized (associated) Laguerre polynomials \cite{bateman}, and $\psi_{x_1}(x_1)$ are eigenfunctions of one-dimensional Schr\"odinger equation with potential $\biggl(P^2(x_1)+P'(x_1)\biggr).$ For the present illustrative task, one can take an arbitrary exactly solvable one-dimensional model such as harmonic oscillator, Morse and P\"oschl-Teller potentials (see the list in the table 1 of \cite{dabrowska}). In turn, the wave functions of the Schr\"odinger equation with three-dimensional partner Hamiltonian $H_2$ can be obtained by action of the supercharge component $Q^-=-\partial_{x_1}+P(x_1)$, i.e. $\Psi^{(2)}_{n,l,m}=Q^-\Psi^{(1)}_{n,l,m},$ and the spectra of superpartners $H_1, \, H_2$ coincide up to zero modes of $Q^{\pm}.$

{\bf Case II.}  $\qquad\qquad a=c=0,\, b\neq 0.$\\
It is evident that the choices $b=c=0,\, a\neq 0$ and $a=b=0,\, c\neq 0$ are analogous. Let us start from solution of (\ref{3-2}):
$$
A=-mbx_ix_i-mex_1+(-ne+2be_3)x_2+(-ke-2be_2)x_3-(me_1+ne_2+ke_3)=0, \nonumber
$$
which means that
$$
m=0;\quad e_3=\frac{ne}{2b};\quad e_2=-\frac{ke}{2b}.  \nonumber
$$
By means of suitable shift of coordinates $x_i,$ the quadratic polynomials $q_i$
can be transformed to the form:
\be
q_1=b(x_1^2-x_2^2-x_3^2)+e_1;\quad q_2=2bx_1x_2; \quad q_3=2bx_1x_3. \label{13-00}
\ee
Now Eqs.(\ref{22}) allow to find restrictions on the form of function $p(\vec x):$
\be
p=p(x_1, x_2^2+x_3^2)\equiv bx_1+P(x_1, x_2^2+x_3^2), \label{15-star}
\ee
with the condition:
\be
\biggl[2x_1x_2\partial_1-(x_1^2-x_2^2-x_3^2+\alpha )\partial_2\biggr]P(x_1, x_2^2+x_3^2)=0;\quad \alpha\equiv\frac{e_1}{b}. \label{15-1}
\ee
In terms of polar coordinates $(\rho, \phi )$ in the plane $(x_2=\rho\cos\phi, x_3=\rho\sin\phi),$ equation (\ref{15-1})
$$
\biggl[2\rho x_1\partial_1-(x_1^2-\rho^2+\alpha )\partial_{\rho} \biggr]P(x_1, \rho)=0 \nonumber
$$
can be solved in terms of arbitrary function $P(Y)$:
$$
P(x_1, \rho)=P(Y); \quad Y\equiv x_1+\frac{\rho^2-\alpha}{x_1}=\frac{r^2-\alpha}{x_1};\quad r^2\equiv (x_1^2+\rho^2). \nonumber
$$
Now, Eq.(\ref{22}) allows to find also the product:
$$
Mv=\frac{1}{bx_1^2}P^{\prime}(Y).   \nonumber
$$
The mass function $M$ can be calculated separately from Eq.(\ref{MM})
$$
\biggl[(x_1^2-x_2^2-x_3^2+\alpha)\partial_1+2x_1(x_1\partial_2+x_3\partial_3) \biggr]M(\vec x)=-4x_1M(\vec x), \nonumber
$$
which being written in terms of special variables $z, \rho, \phi$ with $z\equiv \rho +(x_1^2+\alpha )/\rho$ is simplified essentially
\be
\rho\partial_{\rho}M(z, \rho, \phi)=-2M(z, \rho, \phi), \qquad i.e.\quad M(z, \rho, \phi)=\frac{\widetilde{M}(z, \phi)}{\rho^2}. \label{M}
\ee
Taking into account (\ref{15-star}), the last equation (\ref{1-9}) has the form:
$$
q_i\partial_i(v_2+v)=2vP(Y), \nonumber
$$
and can be solved:
$$
v_2+v=\frac{P^2(Y)}{(z^2-4\alpha)\widetilde{M}(z, \phi)}+L(z, \phi)=\frac{P^2(Y)}{(Y^2+4\alpha)x_1^2 M(z, \rho, \phi)}+L(z, \phi) , \nonumber
$$
with an arbitrary new function $L(z, \phi).$ The expressions for potentials $v_{1, 2}$ are:
\be
v_{1,2}=\frac{1}{x_1^2M(z, \rho, \phi)}\biggl(\pm\frac{P^{\prime}(Y)}{b}+\frac{P^2(Y)}{Y^2+4\alpha} \biggr)+L(z, \phi). \label{c0}
\ee

{\bf Case III.} $\qquad\qquad a=0,\,\,\, bc\neq 0.$\\
According to Eq.(\ref{3-2}), the following relations between constant parameters must be fulfilled:
$$
n=-\frac{mb}{c};\quad e_3=-\frac{me}{2c};\quad e_1=\frac{2be_2+ek}{2c}.   \nonumber
$$
After the suitable constant shifts of coordinates $x_i,$ the functions $q_i$ take the form:
$$
q_1=b(x_1^2-x_2^2-x_3^2)+2cx_1x_2+e_1;\,\, q_2=c(x_2^2-x_1^2-x_3^2)+2bx_1x_2+e_2;\,\, q_3=2cx_2x_3+2bx_1x_3, \nonumber
$$
where $e_1c=e_2b.$ In this case, the convenient variables:
$$y_1=cx_2+bx_1;\quad y_2=bx_2-cx_1;\quad y_3=\sqrt{b^2+c^2}x_3$$
provide the first terms $q_i(\vec x)\partial_i\equiv q_i(\vec y)\partial_{y_i}$ in intertwining operators (\ref{similarity}) with coefficients $q_i(\vec y)$ of the following form:
$$
q_1(\vec y)=y_1^2-y_2^2-y_3^2+e;\quad q_2(\vec y)=2y_1y_2;\quad q_3(\vec y)=2y_1y_3. \nonumber
$$
Thus, this situation coincides with the particular variant of Case II - see (\ref{13-00}) with $b=1.$

{\bf Case IV.} $\qquad\qquad b\equiv a_1; \, c\equiv a_2; \,a\equiv a_3; \, m\equiv b_1; \,n\equiv b_2; \,k\equiv b_3,$\\
with all listed parameters being nonzero. In this case, functions $q_i$ from (\ref{1-4})-(\ref{1-6}) can be written in a compact form
using scalar and vector products:
$$
\vec q=-\vec a r^2+2(\vec a\vec x)\vec x+e\vec x-[\vec b\times\vec x]+\vec e;\quad r^2\equiv \vec x^2. \nonumber
$$
We shall demonstrate that this case can be reduced to the previous Case III (and therefore, to the Case II).

Let us introduce new variables $\vec y,$ which are orthogonal transform of $\vec x$ by $3\times 3$ matrix $A:$
$$
\vec x=A\vec y;\quad \vec y=A^{T}\vec x;\quad AA^{T}=A^{T}A=I;\quad (\vec x)^2=(\vec y)^2;\quad \vec\partial=A\vec\partial_{y}. \nonumber
$$
Correspondingly, in terms of $\vec y$ the first term of the intertwining operator is
$$q_i(\vec x)\partial_i= (\vec q(\vec y)A)_i\vec\partial_{y_i}.$$
In particular, the coefficient in front of $\partial_{y_1}$ is:
\ba
(\vec q(\vec y)A)_1&=&-(\vec a A)_1(y_1^2+y_2^2+y_3^2)+2(\vec aA\vec y)y_1+ linear\,\, terms =\nonumber\\
&=&-(\vec a A)_1(y_1^2-y_2^2-y_3^2)+2(\vec aA)_2y_2y_1+2(\vec aA)_3y_3y_1+ linear\,\, terms. \label{rotation}
\ea
It is clear that for arbitrary vector $\vec a$, one can find such rotations $A$ which make the projection $(\vec aA)_1$ vanishing.
For corresponding variables $\vec y,$ the first term in (\ref{rotation}) disappears, and therefore the considered Case IV in
coordinates $\vec y$ is reduced to the Case III.

\section{\bf Conclusions.}
\vspace*{0.1cm}
\hspace*{3ex}
In this paper, the three-dimensional Schr\"odinger equation with an effective mass was studied in the framework of SUSY Quantum Mechanics.
The general solution of SUSY intertwining relations with first order supercharges was obtained without any preliminary constraints.
Several distinct cases (Ia, Ib.1, Ib.2 and II) depending on the form of coefficient functions $q_i$ of the supercharges were investigated.
It was shown that the two more options (III and IV) can be reduced to the previous ones. As a result, the analytical expressions were obtained for the mass function and
superpartner potentials. The variety of solutions of intertwining relations is rather rich and depends on arbitrary constant parameters and free functions.
As usual for SUSY Quantum Mechanics with nonsingular superpotential, the spectra of intertwined Hamiltonians with potentials $v_1,\, v_2$ coincide up to zero modes of the operators $q^{\pm},$ and the corresponding wave functions are connected according to (\ref{schr}). It should be recalled also that all systems in the present paper are partially integrable by construction: each model with potential $v_{1}$ (or $v_{2}$) has at least one symmetry operator $R_{1}$ (or $R_{2}$) of second order in momenta (see more detail in Section 2).

It might be interesting to compare some of obtained solutions of the intertwining relations with the PDM models found in paper \cite{nikitin}. In contrast to our paper, the main task of \cite{nikitin} was to find rotationally invariant superintegrable PDM models without any use of intertwining relations (the shape invariance was used there only in the context of one-dimensional radial equation). By this reason, one may expect only a partial overlapping of solutions in \cite{nikitin} and the present paper. Nevertheless, let us choose $\widetilde M(z, \phi)\equiv 1/z^2$ in (\ref{M}) of Case II so that the mass function is rotationally invariant: $M=1/(r^2+\alpha )^2.$ Due to the freedom in choice of the function $P(Y)$ in (\ref{c0}), one can fix the expression as a constant:
\be
\biggl(+\frac{P'(Y)}{b}+\frac{P^2(Y)}{Y^2+4\alpha}\biggr)Y^2\equiv\gamma = const. \label{P}
\ee
Substituting (\ref{P}) into (\ref{c0}) with $L(z,\phi)\equiv 0,$ one obtains the potential function $v_1,$ which according to (\ref{vU})
provides the corresponding effective potential $U_1:$
\be
U_1=v_1-6r^2-6=\gamma\frac{(r^2+\alpha )^2}{(r^2-\alpha )^2}-6r^2-6=\frac{4\gamma\alpha r^2}{(r^2-\alpha )^2}-6r^2+ const. \label{UU}
\ee
This expression can be compared with some of effective potentials found in \cite{nikitin}, where they are designated as $V.$ Depending on the choice of $\alpha =0,\,-1,\,+1$
above, (\ref{UU}) coincides with the cases 2, 3, or 4 of the Table 2 in \cite{nikitin}.

For the completeness, we shall mention few papers \cite{cpt-1}, \cite{cpt-2}, \cite{cpt-3} where the multidimensional Schr\"odinger equation with PDM was studied beyond the SUSY QM approach. In these papers, the main tool was the point canonical transformation of coordinates which allowed to link the problem with a suitable standard Schr\"odinger equation with constant mass and rotationally invariant potential. Recently, the relation between two different Schr\"odinger equations: one-dimensional (i.e. in the flat space) with PDM on one side and two-dimensional with constant mass in a curve space on the other was observed in \cite{plyu}. It would be interesting to generalize this relation to the present $d=3$ flat case.

\section{\bf Acknowledgments}

The authors are grateful to the anonymous referees for useful comments and suggestions.
E.V.K. acknowledges Saint-Petersburg State University for a research grant N 11.38.223.2015. 

{}

\end{document}